\documentstyle[12pt]{article}

\newcommand{\beq}{\begin{equation}}
\newcommand{\eeq}{\end{equation}}
\newcommand{\bea}{\begin{eqnarray}}
\newcommand{\eea}{\end{eqnarray}}
\newcommand{\id}{i\!\!\not\!\partial}
\newcommand{\intpp}{\int{\cal D}\bar\psi{\cal D}\psi}

\newcommand{\intf}{\int{\cal D}\phi}

\newcommand{\bs}{\not\! b}
\newcommand{\dslash}{\not\!\partial}
\newcommand{\Ks}{\not\!\! K}

\newcommand{\qs}{/\kern-.52em s}
\newcommand{\ks}{\not\! k}

\begin{document}
\title
{An Alternative Dimensional Reduction Prescription}
\author{J.D. Edelstein\thanks{CONICET, Argentina} ~$^a$,
J.J. Giambiagi $^b$, C. N\'u\~nez $^a$ \\
and F.A. Schaposnik\thanks{Investigador CICBA, Argentina} ~$^a$\\
{\normalsize \it $^a$ Departamento de
F\'\i sica, Universidad Nacional de La Plata}\\
{\normalsize \it C.C. 67, (1900) La Plata, Argentina.}\\
{\normalsize \it and }\\
{\normalsize \it $^b$ Centro Brasileiro de Pesquisas F\'\i sicas,
- CBPF/CNPq~}\\
{\normalsize \it Rua Dr.Xavier Sigaud, 150, 22290 Rio de Janeiro, RJ,
Brasil}\\}

\date{\today}

\maketitle

\def\thepage{\protect\raisebox{0ex}{\ } La Plata 95-22}
\thispagestyle{headings}
\markright{\thepage}

\begin{abstract}
\normalsize
We propose an alternative dimensional reduction prescription which
in respect with Green functions corresponds to drop the extra
spatial coordinate. From this, we construct the dimensionally
reduced Lagrangians both for scalars and fermions, discussing
bosonization and supersymmetry in the particular $2$-dimensional
case.  We argue that our proposal is in some situations more physical
in the sense that it mantains
the form of the interactions between particles thus
preserving the dynamics
correspondig to the  higher dimensional space.

\end{abstract}

\newpage
\pagenumbering{arabic}

Dimensional reduction is a well-honoured procedure to
build  theories in a given number of
space-time dimensions starting from higher dimensional theories.
Initiated with the  Kaluza-Klein proposal
for unifying electromagnetic and gravitational
forces, there are
many  fields where it can be fruitfully exploited.
Let us mention for example its wide application in Supersymmetry and
Supergravity for constructing extended
supersymmetric theories, studying spontaneous breaking of supersymmetry,
etc \cite{SS}. Also in condensed matter problems where charged
particles are constrained to move on a plane, the connection between
the  models in the higher ($3+1$) and lower
($2+1$) dimensional spaces is of relevance \cite{Fradkin}-\cite{Marino1}.

The dimensional reduction procedure, as it is usually applied, consists in
dropping out extra coordinates at the Lagrangian level,
this amounting to mantain  unchanged the differential (kinetic
energy) operator appearing in the Lagrangian.  As a result,
the Green function
in momentum space preserves its form but in configuration space
drastically changes.
For example, if in the higher
dimensional space the kinetic
energy operator is a D'Alembertian, the  same operator will appear
in the lower dimensional theory. Evidently, the D'Alembertian
Green function is different for different number of space-time
dimensions.
This, in a sense to be clarified
below, implies that
the dynamics of the interacting particles described
by the dimensionally reduced Lagrangian
is changed in configuration space.

The statement above can be clarified with an example
which is relevant for
the study of the Quantum Hall effect : electrodynamics
of planar systems (see for example \cite{Fradkin}). If one decides
that electrons
compelled to move in a plane are to be described by a $(2+1)$
gauge theory coupled to matter then, the resulting Coulomb potential
is logarithmic instead of the $1/r$ potential to which electrons are
actually subject even if they move on a plane. In
passing from $3$ to $2$ spatial dimensions dynamics has changed.

The opposite attitude for implementing a dimensional
reduction can also be thought of. Indeed, one can
preserve the form of the Green function in
configuration space
dropping its extra space coordinate dependence.  In the case of electrons
in the plane, their interaction will then still be $1/r$ (with $r$ depending
only on planar coordinates). In this way
one keeps
the dynamics of the higher dimensional space.
As we shall see below, for the case of planar electrons the
Coulomb potential restricted to the plane, $1/\sqrt{x_1^2 + x_2^2}$,
 is the Green function of
the operator $(-\Delta)^{1/2}$:
\beq
(-\Delta)^{1/2} \frac{1}{\sqrt{x_1^2 + x_2^2}} =  \delta^{(2)}(x)
\label{cero}
\eeq
The previous expression can be interpreted using Riesz method of analytical
continuation \cite{Riesz} and distribution theory \cite{Gelfand}
for defining derivation of fractional order.
Moreover, the pseudodifferential
operator $(-\Delta)^{1/2}$ can be properly defined using the
results of Seeley
\cite{Seeley} on complex powers of elliptic operators.
\vspace{0.2cm}

We investigate in the present work  this alternative dimensional
reduction prescription which preserves higher dimensional dynamics in
the restricted lower dimensional space-time manifold.
As we shall see, the prescription
changes the Green function in momentum space and leads
to a change in the kinetic energy operator appearing in the Lagrangian.
 We analyse both
the cases of scalar and fermionic theories and, in particular, we discuss
bosonization and
supersymmetry for the $3 \to 2$ dimensional reduction case.
\vspace{ 0.3 cm}

\noindent {\underline{Scalars}}
\vspace{0.2 cm}

We start with a simple example. Consider the
action for a
scalar field $\phi$ in ($D+1$) Minkowski space-time
dimensions
\beq
S^{(D+1)} = \frac{1}{2}\int d^{D+1}x \phi \Box \phi + S_{int}
\label{auno}
\eeq
where $S_{int}$ includes interactions. The corresponding Green
function is defined by
\beq
\Box G^{(D+1)}(x) = \delta(x)
\label{ados}
\eeq
so that in momentum space one has
\beq
\tilde {G}^{(D+1)}(k) = \frac{1}{ k_0^2 - k_1^2 - \ldots - k_D^2  -i0}
\label{atres}
\eeq
In configuration space one has \cite{BG1}
\begin{eqnarray}
G^{(D+1)}(x) & = & -(i)^{D}2^{D-1}\pi^{(D+1)/2}\Gamma
\left[ \frac{D-1}{2} \right] \times
\nonumber\\
& & \left(t^2 - ({(x^1)}^2 + \ldots + {(x^D)}^2) + i0 \right)^{\frac{1-D}{2}}
\label{acuatro}
\end{eqnarray}

Let us first review how the usual dimensional reduction procedure
manifests at the Green function level and then present our
alternative prescription which, as we shall see, describes
different Physics.

The usual way in which dimensional reduction is
implemented corresponds, in this context,  to
drop $k_D^2$ in eq.(\ref{atres}). The Fourier transform, giving
the configuration space Green function in one dimension less, will be
\begin{eqnarray}
 G^{(d+1)}(x) & = & -(i)^{d}2^{d-1}\pi^{(d+1)/2}
\Gamma \left[ \frac{d-1}{2} \right] \times
\nonumber\\
& &
\left( t^2 - ({(x^1)}^2 + \ldots + {(x^d)}^2) + i0 \right)^{\frac{1-d}{2}}
\label{accuatro}
\end{eqnarray}
\beq
d = D-1
\label{1}
\eeq
Of course, this expression is different from the one we would have obtained
 just by dropping the extra space coordinate
$x_D$ in (\ref{acuatro}).

We are now ready to specify our dimensional reduction prescription:
one  starts from
eq.(\ref{acuatro}) and drops the extra coordinate in the higher
dimensional space Green function $G^{(D+1)}$. The resulting
Green function ${\cal G}^{(d+1)}$  in the dimensionally reduced
space ($d= D-1$) is then defined as
\beq
{\cal G}^{(d+1)}(x) =  -(i)^{d+1}2^{d}\pi^{(d+2)/2}\Gamma[\frac{d}{2}]
\left( t^2 - (x_1^2 + \ldots + x_d^2) + i0
\right)^{-\frac{d}{2}}
\label{acinco}
\eeq
The Fourier transform is given by
\beq
{\tilde {\cal G}}^{(d+1)}(k) =
-2\pi i \frac{\Gamma[\frac{d}{2}]}{\Gamma[\frac{d-1}{2}]}
 \left( k_0^2 - (k_1^2 + \ldots + k_d^2) \right)^{-\frac{1}{2}}
\label{asis}
\eeq
with the appropriate condition for the pole.
One can convince oneself that ${\cal G}^{(d+1)}$ is nothing but the
Green function  for the operator $\Box^{1/2}$ in $d+1$ dimensional
space-time,
\beq
\Box^{1/2}  {\cal G}^{(d+1)}(x) = \delta^{(d+1)}(x)
\label{claro}
\eeq
Then, the action for a scalar theory in the dimensionally reduced space
leading to this Green function is
\beq
S^{(d+1)} = \frac{1}{2}\int d^{d+1}x \phi \Box^{1/2} \phi
\label{adoce}
\eeq
In contrast, had we followed the habitual procedure consisting in
dropping $k_D^2$ in eq.(\ref{atres}), we would had
arrived to the Green function for the operator $\Box$ in $d+1$ dimensions and
the action would correspond to the usual one, given by eq.(\ref{auno}) in
the reduced space.

Let us end this discussion by analysing a  $(3+1)$ case which, as
mentionned above, is relevant for example in the study of electrons
compelled to move in a plane. The retarded Green function for the
$(3+1)$ dimensional classical
system can be obtained from the general formula (\ref{acuatro})
as explained in ref.\cite{Giambia}
\beq
G^{(3+1)}(t,R) = \frac{1}{4\pi R} \delta(t-R)
\label{asiete}
\eeq
where
\beq
R^2 = x_1^2 + x_2^2 + x_3^2
\label{aocho}
\eeq
Its Fourier transform is
\beq
{\tilde{G}}^{(3+1)} = \frac{1}{k_0^2 - K^2}
\label{four}
\eeq
where
\beq
K^2 = k_1^2 + k_2^2 + k_3^2
\label{fouri}
\eeq
If, following our prescription, we drop in eq.(\ref{asiete}) the $x_3$
coordinate, we still have
for the wave equation Green function in the reduced $(2+1)$ space-time
\beq
{\cal G}^{(2+1)} = \frac{1}{4\pi r} \delta(t-r)
\label{anueve}
\eeq
where
\beq
r^2 = x_1^2 + x_2^2
\label{adiez}
\eeq
One can easily see that ${\cal G}^{(2+1)}$ satisfies
\beq
\Box ^{1/2} {\cal G}^{(2+1)}(x) = \delta^{(3)}(x)
\label{aonce}
\eeq

Hence, there are two ways of doing dimensional reduction. The
usual one, looked upon from the point of view of Green
functions, is the one which drops the extra momentum space coordinate.
The one we propose in the present work consists in dropping
the extra spacial coordinate in the Green function and, from it,
infer the resulting dimensionally reduced Lagrangian. This last
prescription seems to be more physical in situations as that
described for planar electrons which are suppose to be subject
to $(3+1)$ interactions even when they are compelled to move in the
plane.

It is important to stress that, independently of the number of dimensions,
an action with the operator $\Box^{1/2}$
is always obtained if one starts from action (\ref{auno})
when one applies the dimensional reduction prescription advocated
here. In summary,  when one passes from $D$  to $D-1 = d$
spacial dimensions, the action changes as follows

\beq
\frac{1}{2}\int d^{D+1}x \phi \Box \phi ~~ \stackrel{D \to d}
{\longrightarrow} ~~
\frac{1}{2}\int d^{d+1}x \phi \Box^{1/2} \phi
\label{dif}
\eeq
\vspace{ 0.3 cm}

\noindent {\underline{Fermions}}
\vspace{0.2 cm}

Eq.(\ref{dif}) gives the rule for the
change in the scalar field action  under our dimensional
reduction prescription. The case of fermions can be treated analogously. As
an example we will
describe here reduction from ($2+1$) to ($1+1$) dimensional space-times
and then analyse how bosonization works for the reduced
two-dimensional fermionic theory.

We start from  (two-component) free Dirac fermions in Euclidean
$3$ dimensional space-time and take for the Dirac matrices
 $\gamma_0 = \sigma^1$, $\gamma_1 = \sigma^2$ and
$\gamma_2 = \sigma^3$ with $\sigma^a$ the Pauli matrices.
To obtain the dimensionally reduced fermion
action,  we consider the fermion Green function,
\beq
G^{(3)}(X) = -\int \frac{d^3K}{(2\pi)^3} \exp(i K X) \frac{\Ks}{K^2}
\label{seiss}
\eeq
We use $X$ (in general capital letters) for
variables in $3$ dimensional space time ($X = (X^0,X^1,X^2$)).
If we make  $X^2 = 0$ in (\ref{seiss}) and integrate out over $K_2$
we obtain
\beq
{\cal G}^{(2)}(x) \equiv G^{(3)}(X)\vert_{X^2 = 0} =
-\frac{1}{4\pi} \int \frac{d^2k}{(2 \pi)^2} \exp(i k x) \frac{\ks}{k}
\label{siete}
\eeq
One can easily see that the resulting  Green function ${\cal G}^{(2)}(x)$
in the reduced two-dimensional space satisfies
\beq
\frac{\id}{(-\Box)^{1/2}}{\cal G}^{(2)}(x,x') = \frac{1}{4\pi}
\delta^{(2)}(x-x')
\label{cien1}
\eeq
 From this result  we can infer
the corresponding two-dimensional fermionic
action
\beq
S^{(2)}_{fermion} = \int d^2x \bar \psi \frac{\id}{(-\Box)^{1/2}} \psi
\label{ocho}
\eeq
As in the scalar case, we see that our dimensional reduction prescription
reduces to change the operator $A$ appearing in the original Lagrangian
to ${A}/{(-\Box)^{1/2}}$. In the fermionic case we have arrived
to a non-local expression where the Dirac operator $\id$ appears
convolutionned with the Green function $((-\Box)^{-1/2})_{xy}$.

We shall now investigate how bosonization works when the fermionic action
is given by eq.(\ref{ocho}). To this end, we
consider the two-dimensional partition function
\beq
Z_F = \intpp \exp \left(-\int\bar\psi\frac{\id}{(-\Box)^{1/2}}
\psi d^2x \right)
\label{bos1}
\eeq
and follow the path-integral bosonization approach described in
\cite{FAS}. This approach starts by performing  the
change of variables
\beq
\psi \rightarrow \exp\left({i\theta(x)}\right)\psi
\label{aux1}
\eeq
\beq
\bar \psi \rightarrow \bar \psi  \exp\left({-i\theta(x)}\right)
\label{aux11}
\eeq
with $\theta$ a real function.
After this change, the partition function reads
\beq
Z_F = \intpp \exp{-\int\bar\psi\left(\frac{\id}{(-\Box)^{1/2}} + i
\frac{\id}{(-\Box)^{1/2}}\theta\right)\psi d^2x}
\label{bos2}
\eeq
Being $Z_F$ $\theta$-independent, we can integrate out over $\theta$
both sides in eq.(\ref{bos2}),
this amounting to a trivial change in the normalization of the
path-integral
\beq
Z_F = {\cal N} \intpp {\cal D}\theta
\exp{-\int\bar\psi\left(\frac{\id}{\Box^{1/2}} +
\dslash \alpha\right)\psi d^2x}
\label{bos22}
\eeq
with
\beq
\alpha (x) \equiv -{(-\Box)^{-1/2}}\theta =
-\int d^2y {\left ( (-\Box)^{-1/2} \right )}_{x y}\theta (y)
\label{alf}
\eeq
It is evident that $\partial_\mu \alpha$ in (\ref{bos22}) can be thougth as
a flat connection and hence it can be replaced by a ``true''
gauge field provided a constraint is introduced to assure its flatness.
Hence, we can replace the $\theta$ integration by an integration
over a flat connection $b_\mu$ by writing
\beq
Z_F = {\cal N} \intpp Db_\mu \delta (\epsilon_{\mu \nu} f_{\mu \nu})
\exp{-\int\bar\psi\left(\frac{\id}{(-\Box)^{1/2}} +  \bs \right)\psi d^2x}
\label{bos3}
\eeq
where
\beq
f_{\mu\nu} = \partial_{\mu}b_{\nu} - \partial_{\nu}b_{\mu}
\label{aux3}
\eeq
Here $\mu =0,1$ labels  components in the reduced space-time.
Now, performing the fermionic path integral, one has
\beq
Z_F =  \int {\cal D}b_\mu \delta (\epsilon_{\mu \nu} f_{\mu \nu})
{\rm det} \left( \frac{\id}{(-\Box)^{1/2}} +  \bs \right)
\label{det}
\eeq
One can compute the two-dimensional fermionic determinant
in (\ref{det}) as a Fujikawa jacobian following the
method described for the Schwinger model determinant
in ref.\cite{RS}.
The answer is
\beq
\log{\rm det} \left(\frac{\id}{(-\Box)^{1/2}} - \bs\right)
 =  -\frac{1}{2\pi}\int d^2x b_{\mu}
\left( (-\Box)^{-1/2} \delta_{\mu \nu}  - \partial_\mu (-\Box)^{-3/2}
\partial_\nu \right)
b_{\nu}
\label{aux4}
\eeq
This, together with the representation
\beq
\delta[\epsilon_{\mu \nu}f_{\mu \nu}] = \intf \exp(-\frac{1}{\sqrt{\pi}}
\int d^2x \phi \epsilon_{\mu \nu}f_{\mu \nu})
\label{lag2}
\eeq
leads, after a trivial gaussian integration over $b_\mu$ to the result:
\beq
Z_F = \intf \exp \left({-\frac{1}{2\pi}\int d^2x
\phi(-\Box)^{1/2}\phi} \right)
\label{bos8}
\eeq
Hence,  as one should expect, the two-dimensional
fermion action (\ref{ocho}), obtained within our
dimensional reduction prescription,  bosonizes to the two-dimensional
scalar action (eq.(\ref{adoce})), precisely the one we obtained
when applying the prescription to scalar fields.

Concerning bosonization rules for fermion currents, let us
note that the addition of a fermion source
$s_{\mu}$ in $Z_F$ amounts to the inclussion of
this source in the fermion determinant
\beq
Z_{F}[s]= \int {\cal D}b_\mu  {\rm det}
\left( \frac{\id}{(-\Box)^{1/2}} + \bs
+ \qs \right)
\delta[\epsilon_{\mu\nu} f_{\mu \nu}] .
\label{in2}
\eeq
Now, a trivial shift $ b + s \to b$ in the integration variable $b$ puts
the source dependence into the constraint
\beq
Z_{F}[s]= \int {\cal D}b_\mu  {\rm det}( \frac{\id}{(-\Box)^{1/2}} + \bs)
\delta[\epsilon_{\mu\nu} (f_{\mu \nu} - 2\partial_\mu s_\nu)]
\label{in22}
\eeq
so that, instead of (\ref{bos8}) one ends with
\beq
Z_F[s] = \intf \exp (-\frac{1}{2}\int d^2x
(\phi (-\Box)^{1/2} \phi + \frac{2}{\sqrt \pi}s_\mu
\epsilon_{\mu \nu} \partial_\nu \phi)) .
\label{Z2}
\eeq
By simple differentiation with respect to the source
one infers from this expression the bosonization recipe for
$j_\mu$
\beq
\bar \psi \gamma^{\mu} \psi  \to {({1}/{\sqrt{\pi}})}
\epsilon^{\mu \nu} \partial_\nu \phi
\label{101}
\eeq
which coincides with the usual one \cite{CM} except for the fact
that the partition function with which one has to work in the scalar theory
is given by eq.(\ref{bos8}).

Another interesting issue where our alternative dimensional
prescription can be investigated concerns supersymmetric models.
The simplest supersymmetric action that one can write in $3$-dimensional
space is
\beq
S^{(3)} = \int d^3X ( \phi^* \Box \phi - \bar \psi \id \psi) ~.
\label{suno}
\eeq
Here $\phi$  is a complex scalar and $\psi$  a two component Dirac
fermion.
Action (\ref{suno}) is invariant under the
supersymmetry transformation

\beq
\delta \phi = \bar \epsilon \psi
\label{dos}
\eeq

\beq
\delta \psi = (\id \phi)  \epsilon
\label{tres}
\eeq
Here $\epsilon$ is the real parameter associated
with the supersymmetry transformation.

Using our dimensional reduction prescription one ends with
a two dimensional  action of the form
\beq
S^{(2)} =  \int d^2x \phi \Box^{1/2} \phi
-\int d^2x \bar \psi \frac{\id}{\Box^{1/2}} \psi
\label{nueve}
\eeq
One can prove that this action is invariant under the supersymmetry
transformations (\ref{dos})-(\ref{tres}) now interpreted in two
space-time dimensions.
Hence, as in the case of bosonization, we see that our dimensional
reduction prescription can be consistently applied in the
case of supersymmetric models.
\vspace{2 mm}

In summary, we have presented an alternative way for dimensional
reduction which, looked upon  from the point of view of Green functions,
implies to be attached to the form that they take in configuration
space and just drop extra coordinates. From the resulting
reduced Green function one can infer the form that the Lagrangian
takes in  the dimensionally reduced space-time. This can be
done in arbitrary number of dimensions. For the scalar theory,
the answer is given in eq.(\ref{adoce}). Concerning fermions,
we have discused the $(2+1)$ $\to$ $(1+1)$ case (eq.(\ref{ocho})
giving the reduced action) but more general cases can be
envisaged. Of course, one has to take into account the appropriate
number of spinor components in different number of space-time dimensions
but this can be handled in the same way one does within the
habitual dimensional reduction approach.

A  comment is in order concerning the nonlocality arising from the
$\Box^{-1/2}$ kernel appearing in the dimensionally
reduced Lagrangians. As shown  in \cite{BG1}-\cite{Giambia},
\cite{Giambia1}, by choosing the appropriate retarded or
advanced prescriptions causality is in fact
respected since the kernel $\Box^{-1/2}$ has support in the light-cone
surface.

Our coments  on the $(3+1)$ $\to$ $(2+1)$ case in connection
with electrons compelled to move in the plane
shows that in certain
situations our prescription seems to be more physical
than the usual one in the sense
that it preserves the form of the interaction; the
resulting reduced Lagrangian can be thought to give an effective
description of this interaction. This is precisely the approach
undertaken by Marino in ref.\cite{Marino1} in his study of $QED$
for particles on a plane. In this work, the relation
between the resulting effective gauge theory and the strictly
$(2+1)$ Chern-Simons theory usually employed to describe
fractional statistics is explored and the potential applications
in condensed matter physics are discussed.

We have studied in some detail bosonization of free fermions
in the reduced $(1+1)$ theory showing that bosonization recipe
can be derived and are much the same as for the usual
Dirac Lagrangian. This subject, as well as the issue of
bosonization in $(2+1)$ space-time deserve a more thorough
analysis. We hope to report on this aspects in a future work.
\newpage

\underline{Acknowledgements}:
This work was
supported in part by  CICBA and CONICET, Argentina.
F.A.S. ~ thanks the Centro Brasileiro de Pesquisas
F\'\i sicas for its kind hospitality.


\begin{thebibliography}{99}
\bibitem{SS} See for example Supergravities in diverse dimensions, Vol.2,
eds. A. Salam and E. Sezgin, North-Holland, 1989 and references
therein.
\bibitem{Fradkin} E. Fradkin, Field Theories of Condensed Matter Systems,
Addisson-Wesley, 1991 and references therein.
\bibitem{Marino1} E. Marino, Nucl. Phys. {\bf B408} (1993) 551.
\bibitem{Riesz} M.Riesz, Acta Math. {\bf 81} (1948).
\bibitem{Gelfand} I.M. Gelfand and G.E. Shilov, Les Distributions,
Ed. Dunod, Paris, 1962.
\bibitem{Seeley} R.T. Seeley, Am. Math. Soc. Proc. Symp. Pure Math. {\bf
10}(1967)288.
\bibitem{BG1} C.G. Bollini and J.J. Giambiagi, Journal of Mathematical
Physics {\bf 34} (1993) 2.
\bibitem{Giambia} J.J. Giambiagi,  Nuovo Cimento {\bf 109B} (1994) 635.
\bibitem{FAS} F.A. Schaposnik, Physics Letters in press, hep-th/95
\bibitem{RS} R. Roskies and F.A. Schaposnik, Phys. Rev. {\bf D23} (1981) 558.
\bibitem{CM}S. Coleman, Phys. Rev. {\bf D 11} (1975) 2088;\hfill\break
S. Mandelstam, Phys. Rev. {\bf D 11} (1975) 3026.
\bibitem{Giambia1} J.J. Giambiagi, Nuovo Cimento {\bf 104A} (1991) 1841.

\end{thebibliography}
\end{document}